# Secure Authentication of Cloud Data Mining API

Rohit Bhadauria*
Department of Electrical and
Computer Engineering,
National University of Singapore,
bhadauria.rohit@gmail.com

Rajdeep Borgohain
Department of Computer
Science and Engineering,
Dibrugarh University, India
rajdeepgohain@gmail.com

Abirlal Biswas
Perl Datamine Team
Wizards Technologies
Pvt. Ltd.
Bangalore, India
abirlalbiswas@gmail.com

Sugata Sanyal
Corporate Technology Office
Tata Consultancy Services
Mumbai, India
sugata.sanyal@tcs.com

*Corresponding Author

*Abstract—* *Cloud computing is a revolutionary concept that has brought a paradigm shift in the IT world. This has made it possible to manage and run businesses without even setting up an IT infrastructure. It offers multifold benefits to the users moving to a cloud, while posing unknown security and privacy issues. User authentication is one such growing concern and is greatly needed in order to ensure privacy and security in a cloud computing environment. This paper discusses the security at different levels viz. network, application and virtualization, in a cloud computing environment. A security framework based on one-time pass key mechanism has been proposed. The uniqueness of the proposed security protocol lies in the fact, that it provides security to both the service providers as well the users in a highly conflicting cloud environment.*

## I. INTRODUCTION

Internet has given rise to one of the most revolutionary concepts of recent times, known as Cloud Computing.

The Cloud, as it is often referred to, involves using computing resources – hardware and software – that are delivered as a service over the Internet. Organizations are no longer required to build their own IT infrastructure. Instead, they are presented with an alternative to host their data on a third party system such that they would be able to access the same by means of Internet. Cloud computing is gaining popularity due to the features that include scalability, multi-tenancy and reduced hardware and maintenance cost. Cloud technologies are enabling the users with multifold facilities but at the same time, they bring additional security and privacy issues. The rate at which cloud technologies are being adopted, it has become imperative to analyze the service offerings from different cloud service providers (CSPs) and then decide their suitability based on the organization's needs and requirements.

Data mining technology and services refer to yet another interesting domain that has caught the attention of researchers in the recent times. It is the process of analyzing data from different perspectives and summarizing it into useful information. It finds a great deal of use in business and economics.

The emerging usage of cloud computing trends provides its users with the unique benefit of unprecedented access to valuable data. This will enable the users to gain valuable insights towards achieving their business goals intelligently.

Simple Cloud API project focuses on improving the portability of PHP based applications across different cloud platforms facilitates services offered by different cloud service providers to be accessed through a common API [31].

Now-a-days, most of the CLOUD API systems use passkey based authentication methods. Passkeys are very popular and useful as they are convenient for users and easy to implement. Passkey-based authentication, although very convenient, have some drawbacks due to the nature of the system.

It has been a tendency to choose relatively simpler ways to implement passkey authentication system. This in-turn makes it less secure and vulnerable and hence susceptible to exhaustive use or attacks. There are several examples of this type of attacks on various systems worldwide. Some of the relatively secure authentication techniques include ZIGSAW based secure data transfer [8] and encryption keys based on RSA technique.

Another approach to design a security platform involves making use of one-way hash functions. They act as the building blocks of a security system that attempts to eliminate online dictionary attacks by implementing a "challenge-response system"[5]. This challenge-response system is designed in a way that does not pose any difficulty to a real user, but is time and computationally intensive for an adversary trying to launch a large number of login requests per unit time as in the case of an online dictionary attack. This system is stateless and therefore attacks are less vulnerable to DoS (Denial of Service) [12] [14]. Also, it is quite tough to implement Intrusion Detection System (IDS) [9][10][11] based protection approach. So, passkey based approach becomes significant in these type of systems.

But this approach is unable to assure a fully secure way for CLOUD API access. Security is a major concern in public clouds than in an internal environment. User has no control and information of any other code sitting on the same machine. Unaware of this, user may allow public access to the API leading to a security breach. Thus we see that security remains an ongoing concern in cloud deployments [6][42].

## II. OVERVIEW

Data mining techniques and applications are needed in a cloud computing paradigm. Cloud based technologies are finding a great deal of use in the fields related to business and scientific computing. Data mining and warehousing techniques targeted to applications such as: fraud detection, prediction of potential threats, identification of criminal suspects etc. are being used in cloud computing scenarios.

### A. Cloud Computing

Cloud computing can be defined as one of the most popular trends in the history of online computing that has taken the world by a surprise. It offers a flexible IT architecture, enabling its users to be able to use services which would have been considered impossible in case of standard IT based solutions.

A cloud based architecture can be defined as a set of resources - hardware and software, which combine together to deliver the aspects of computing as a service. Services in such a scenario are charged on a usage based pricing model and the users are no longer required to care about the intricacies which are needed to be taken care in a traditional on-premise computing model.

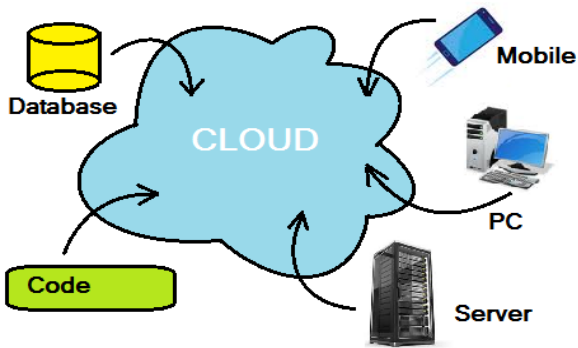

**Figure1. Cloud Model**

Cloud based service models can be categorized into three types: Infrastructure as a Service (IaaS) [18], Platform as a Service (PaaS) [19][20][21], and Software as a Service (SaaS)[17] [27]. Managing a cloud computing service level via the surrounding management layers are as follows.

- **Infrastructure as a Service (IaaS).** The IaaS [22][23] [24][25][26] layer offers storage and computing resources that are used by the developers and IT organizations to deliver high valued business solutions. The core of IaaS is based on virtualization. In an IaaS model, resources can easily be scaled up, depending upon the demand from the user, services being charged in a pay-per use model.
- **Software as a Service (SaaS).** In the SaaS[28][29] layer, the service provider hosts the software such that the user has no need to install it, manage it, or buy hardware for it. Users are only required to connect and make use of the services provided. SaaS examples include high valued customer relationship management as a service.
- **Platform as a Service (PaaS).** The idea behind PaaS is to provide a platform that would enable the developers to build applications and perform end to end testing. These applications may then be deployed on a web-based model such that they can be refined if a need arises.

A few state of the art techniques that distinguish cloud computing from other computing paradigms like grid computing, global computing etc. include elasticity, scalability, self-service provisioning or automatic de-provisioning, application programming interfaces (APIs) [13], and billing and metering of service usage in a "pay-as-you-go" model. These flexibilities are attracting individuals as well as the businesses to move to better suited cloud platforms rather than managing the whole IT infrastructure by themselves.

The three types of users that are predominantly seen in a cloud computing environment are:

- The end user who needn't know anything about the underlying technology and the architecture.
- Business management who needs to take responsibility for the governance of data and services lying in a cloud. Cloud service providers must provide a predictable and guaranteed authenticated service level and security to all their users and constituents.
- The cloud service providers, who are also responsible for IT assets and maintenance.

Cloud computing can be deployed in four different ways: public cloud, private cloud, hybrid cloud – combination of both public and private and community cloud.

Cloud computing can significantly change the way how companies are using technologies to serve customers, partners, and suppliers. Organizations have already started leveraging the benefits offered in a cloud environment in the form of cloud- based solutions which include Qualcomm's wireless solutions on the cloud, Oracle's ERP solutions on the cloud, Schneider Electric's energy solutions on the cloud etc.

### B. Data Mining

Data Mining [30] techniques are most often used to analyze data in the fields of finance, supply chain management (SCM), customer relationship management (CRM), marketing and distribution [7]. For instance, it helps in optimizing customer related data, determining the buying potentials of customers and predicting sales figures by the usage of statistical-mathematical methods implemented over large sets of historical data. Thereby companies can make a blueprint of a new and improved marketing strategy, by spending less amount as well as time to achieve better and effective statistical results.

In essence, data mining techniques enable the companies to predict the market condition, customer response and estimate the sales figures for the upcoming period. With data mining, a retailer could use the historical data of customer purchases to set targeted promotions. By mining demographic data from review comments or feedback form, the retailer could launch new promotions and appeal to specific customer segments.

A generalized framework describing the various stages in a data mining algorithm is explained below:

1. Extraction and Pre-processing of raw data: This includes data collection and then implementing techniques to perform dimensionality reduction and redundancy removal.
2. Pattern Discovery and Analysis: Once the data is pre-processed, it undergoes algorithmic analysis to discover if there exists any pattern or similarity among the different classes of processed data. These patterns are then extrapolated to predict the behavior in future, as per the need.
3. Information retrieval and Data prediction: Once a pattern is located, it is then used to retrieve the information or predict the behavior in future. For e.g. estimating the sales in future, stock levels to be maintained in the inventory, predicting the class of data etc.

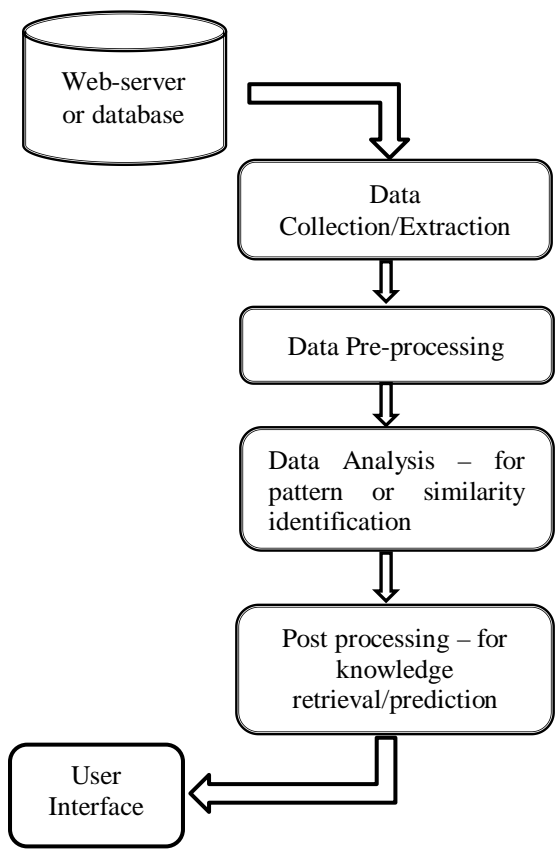

**Figure2. End to End Data Mining Network**

Data mining tasks can be classified under the following sections:

- **Clusters formation**: Data items are populated to organize the grouped segment according to logical relationships or consumers' strategic preferences. For example, data can be populated to identify sales segments or consumers' purchase affinity trends.
- **Classes**: Populated and segregated data is used to indicate the class, data belongs to.
- **Associations**: Data can be populated to identify associations of segments.
- **Sequential patterns**: Data is populated to predict behavioral patterns and trends of segregated data.
- **Regression**: Extrapolating the model (based on the known response) to estimate the values for an unknown segment.

A few of the techniques to implement these algorithms are: support vector machines, genetic algorithms, artificial neural networks, nearest neighbor algorithms etc.

### C. Cloud Data Mining

An increasing trend towards the adoption of cloud services has mandated the need of data mining algorithms in a cloud environment. Data mining in the cloud is the process of extracting useful information from huge chunks of data basically referred to as 'Big Data'. It has given rise to a non-relational database model, better suited to companies operating on a cloud based model. NoSQL is one such cloud-friendly approach that does not follow the standard concepts of a relational database and provides greater scalability and flexibility. Some of the concepts which have been designed in order to ensure smooth data processing in a cloud computing environment include: Apache Hadoop, Apache Hive, Pig, MapReduce etc. A detailed analysis of these techniques has been carried out in [32].

A cloud-based data mining platform which demonstrates the solution of data mining as a service (DMaaS) has been presented in [33]. In this paper, the authors have gone on to propose a comprehensive framework suitable to work with Big Data in a cloud environment. The end user will be able to access the service by means of a light weight browser. Moreover they can also design the analytics flow in a drag and drop manner and interact with the results obtained [33]. Thus we see that data mining services will be offered in the same pay-per-use model as the other application/software based services are provided in the cloud.

### D. Cloud API

Cloud computing is a style of computing in which dynamically scalable and deployable resources are provided as a service over the network. Users need not have knowledge of system, expertise in underlying architecture, or control over the underlying infrastructure. A cloud API is basically used to integrate applications in order to enhance the cloud experience and provide inter-cloud compatibility. They are broadly classified into two categories: in-process APIs and remote APIs. In-process APIs are the ones used on a regular basis and are most commonly used in a typical infrastructure based IT environment. However, remote APIs are the ones which are used to develop cross-border, bridging applications and include web-services (SOAP or REST), remote calls (Sun RPC, Java RMI), application dependent protocols (FTP, SNMP). These types of APIs are based on HTTP and SHTTP protocols and GET, PUT, POST, DELETE requests are used. These types of APIs communicate based on the data structures like: JSON, XML etc. Most of the cloud service providers are found to exploit the second type of API model. A detailed analysis of API requirements and architecture has been carried out in [35].

Companies are now-a-days opting for multiple cloud service providers and the existence of multiple cloud platforms makes cross-platform APIs a necessity. This particular need to enable the companies to have a connected platform, has generated a new cloud API market. Some of the companies operating in this domain are: Google Compute Engine, Citrix, VMware, Simple Cloud, Amazon web services API etc.

*E. Cloud API Security Concern*

If on one hand cloud computing is enabling the organizations to function without bothering about the need to maintain an IT infrastructure, update the software's, host a team of database administrators, it is posing many questions that need to be answered, concerning the security and privacy in a cloud computing environment. The organizations have already started adopting the cloud offerings and the days are not far when a huge chunk of organizational data will reside on cloud-based servers. This will require stringent security measures to be placed across the cloud platforms which will safeguard the data from internal as well as external security threats. Some of the measures that could be taken in this regard are:
1. Ensuring proper security measures to safeguard hypervisor to any sort of security threat.
2. Careful assessment of the security practices as implemented by the cloud service providers need to be done before adopting any of them
3. Proper SLAs between the customer and the CSP, defining the organizations' security requirements that need to be addressed.
4. APIs in use need to be looked after and screened carefully. In the current scenario, most of the organizations prefer an integration of security techniques with their service models. They should be aware of the security implications associated with the usage of these cloud services. Reliance on weak APIs may jeopardize the security of important organizational data.

"Insecure APIs" constitute one of the major security concerns in cloud computing. APIs are defined as a set of interfaces which are used to interact with the cloud services in a cloud based model. These APIs are most often used by the cloud service providers (CSPs) to offer additional services. A two stage API access control mechanism using the Role Based Access Control Model has been proposed in [34]. This model is based on providing access to the users based on the roles assigned to them. Each user is assigned a role and it serves as a connecting layer between the user and the permissions assigned to him/her. This however doesn't guarantee protection against the threats arising from within the organization such as: mishandling of resources from an authorized user (granted a role with high level of responsibilities assigned to it). Hence it is required to carry out the inspection of roles and responsibilities assigned, at regular intervals.

The need for security and privacy in today's highly conflicting computing world has provided cloud service providers with an opportunity to provide security solutions. Numerous cloud service providers are making use of cloud-based models to deliver these services. However, this makes it necessary for the organizations to evaluate the service offerings of different cloud service providers before adopting them. They need to understand the basic principles of operation in a cloud environment and use this knowledge to identify the security features needed to ensure security against possible external and internal security threats.

### III. SECURITY IN THE CLOUD

Security in a cloud environment can be broadly classified into three categories: Network level security, Application level security and Virtualization security. It has been observed at multiple instances that the active cloud virtual machine instances are accessible through public cloud and allow hackers to leverage this opportunity to carry out DoS attacks. Cloud instances running on public cloud are most prone to these types of attacks and hence require a network level access control solution that would enable the delivery of cloud services in a highly protected environment. A detailed analysis of network level attacks has been carried out in [39]. The authors have gone on to propose a network based access control solution that provides additional security against such types of attacks. A few of the security threats that could be classified as network level attacks include: DoS attacks, DDoS attacks, Sniffer attacks, BGP prefix hijacking, DNS attacks, Man in the Middle attacks etc. A detailed analysis of these types of attacks has been carried out in [6].

Application level security refers to securing applications from any type of security attack in a cloud computing environment. Application security is important in the sense that it can be exploited to extract sensitive information or nefariously used to make inappropriate changes to important data. An evaluation model that can be used to assess the risks in moving a service to the cloud has been presented in [41]. The authors have focused on integrating end to end services in a secure manner in a cloud computing model.

Virtualization security refers to securing a VM or a hypervisor in a highly virtualized and distributed cloud environment. In a virtualized environment, hypervisor is defined as a virtual machine monitor that allows many VMs to be deployed on a single OS or multiple operating systems to run on a system at the same time. In brief, hypervisor can be referred to as a controller that monitors/controls the activities of all virtual/guest machines operating in a virtualized environment. There are risks associated with using the same physical infrastructure and even a small number of malicious users may cause threats to the others operating in the same environment. Since the VMs are mobile, they may switch between the hypervisors depending upon the availability of resources. These VMs are most often subjected to risks when they are moving. A few of the security threats, that could compromise the functioning of other VMs and of the hypervisor in a virtualized environment include: VM hopping [36], VM escape [36], mobility etc. Virtual machine images can also be exploited by attackers to launch a security attack. A virtual machine image contains information of the installed and configured applications and is used to restore the desired or initial state of the system. These images may be exploited to leak sensitive information. And hence it is necessary to keep them up-to-date with latest security patches [40].

Virtual network vulnerabilities such as: sniffing and spoofing virtual networks have been discussed in detail in [37]. Since the VMs share the same set of resources in a virtual network, it is highly possible to carry out the above mentioned activities. The authors [37] have leveraged the characteristics of different operating modes of a Xen hypervisor to propose a novel virtual network model that would make the communication (between the VMs) more secure and reliable.

Data security applicable to various states of data such as: data-in-transit and data-at-rest need to be considered. Security in case of data-at-rest refers to providing secure storage by making use of encryption techniques. The major limitation of this approach is that the data cannot be processed without decrypting it or without revealing the keys used for encryption [38]. A detailed analysis of techniques that could enable data processing without disclosing the keys has been carried out in [38].

## IV. PROPOSED SECURITY MECHANISM

The secure extraction of useful data mining information via Data Mining and Clustering API depends on a major factor: proper Authenticity and Security. Various mining or clustering API are used for numerous purposes. Some mining or clustering APIs are less secure to unauthorized access of information that violates the data and report privacy.

To make this mechanism secure, two types of security measures are proposed here. First one is the use of "The One Time Password System" [2] as pass key for authentication of API user and second is the implementation of Cloud Service User Authentication Agent (CS_A) [3] at the Server Side to authenticate the API user and client host details i.e. "Domain Trust" [1].

### A. The system mechanism process of "One Time Pass Key":

The one time passkey is remembered by the Server Side API. CS_A (Server Side Authentication Agent) contains the following information:
- The User ID
- A counter p, where p > 0 and which gets decremented every time CS_A authenticates an User ID
- The hash function Hp (k), i.e. $H(H(…(H(k))…))$

Each time the sever side API wants to start access process; it has to choose passkey k and p in order to authenticate with the server side authentication agent CS_A. It then starts p times iterations of the one way hash function over this pass key k, i.e. Hp(k). Server side API user then securely transmits p and Hp(k) along with user-id to CS_A to initialize the system.

For authentication, server side API user sends user-id details with credentials to CS_A which in turn sends p. Then Server side API user computes Hp-1(k) and sends the result to CS_A along with the next "one time pass key (OTPK)" [4]. CS_A starts computing the hash calculation on the received OTPK and compares it with the stored Hp(k). If they match, CS_A overwrites Hp(k) with the received Hp-1(k) and decrements p.

After the login credentials of server side API are authenticated, the attackers cannot determine Hp-1(K) from Hp(K). Since, the hash function cannot be determined, the system is against both eavesdropping and server database compromise. When p reaches 1, after getting decremented for server authentication of user ID, the Server side API user should select a new password and should reinitialize the system. There is no known secure way of automatic re-initialization and it should be done using Server Side API User Handler function.

Here, H is a one-way hash function such as MD5 [16] or RSA [15] and k is the pass key. None of the stored information is considered to be sensitive to security. Therefore the "one time pass word" [2] system is suitable for authentication in this scenarios wherein the Client side API user procedure is considered not to be trusted or is vulnerable to compromise.

The first step is the authentication of domain using "Domain Trust" mechanism. It checks the client IP from where the request is made for Cloud API is trusted or not. If it is found to be "trusted", then CS_A procedure checks if the user id exists or not. If user id exists in the database then it checks its credential. If all the credentials are found correct then only the API is authenticated by CS_A and also executed for mining the data.

## V. CONCLUSION

The proposed approach integrates CLOUD API User IP authentication along with One Time Key based User Authentication by discarding malicious users from the domain reducing unauthorized access of API. It also increases the security overhead using User's ID based on "One Time Pass Key" using Hashing principle and this framework fails to prevent malicious activity using any malicious code or parameter transfer procedure. In future fraudulent activity identifying approach can be added to this proposed approach which in turn makes the system to generate an alert about unauthorized fraudulent activity. Also this would help to prevent unauthorized accesses to cloud data as well as process. The procedure may also be extended to eradicate unauthorized data access and prevention in a heterogeneous cloud computing platform.